\documentclass[conference]{IEEEtran}

\usepackage{cite}
\usepackage{amsmath,amssymb,amsfonts}
\usepackage{algorithmic}
\usepackage{graphicx}
\usepackage{textcomp}
\usepackage{xcolor}
\usepackage{booktabs}
\usepackage{array}

\def\BibTeX{{\rm B\kern-.05em{\sc i\kern-.025em b}\kern-.08em
    T\kern-.1667em\lower.7ex\hbox{E}\kern-.125emX}}

\begin{document}

\title{Critical Thinking in the Age of Artificial Intelligence: A Survey-Based Study with Machine Learning Insights}

\author{
\IEEEauthorblockN{M Murshidul Bari, Akif Islam, Mohd Ruhul Ameen, Abu Saleh Musa Miah, and Jungpil Shin}

\IEEEauthorblockA{University of Rajshahi, Bangladesh}
\IEEEauthorblockA{Marshall University, USA}
\IEEEauthorblockA{University of Aizu, Japan}

\IEEEauthorblockA{bari.murshidul10@gmail.com, iamakifislam@gmail.com, ameen@marshall.edu, musa@u-aizu.ac.jp, jpshin@u-aizu.ac.jp}
}

\maketitle

\begin{abstract}
The growing use of artificial intelligence (AI) in education, professional work, and everyday problem-solving has raised important questions about its effect on human reasoning. While AI can improve efficiency, save time, and support learning, repeated dependence on it may also encourage cognitive offloading, reduce productive struggle, and weaken independent critical thinking. This paper investigates the relationship between AI-use behavior and critical-thinking performance through an interview-based survey combined with short logic and reasoning tasks. The instrument included a behavioral section capturing AI-use habits, perceived confidence, discomfort without AI, patience in problem-solving, and perceived efficiency, as well as a cognitive section designed to assess unaided analytical performance. Responses to the reasoning tasks were converted into a Critical Thinking Score (CTS), while the behavioral variables were used for descriptive and exploratory machine learning analysis. The findings reveal a mixed pattern: participants largely viewed AI as a tool for speed, convenience, and learning support, yet many also reported reduced patience for sustained effort. Objective reasoning performance varied considerably across individuals, and the analyses suggest that reduced patience and stronger dependence-related tendencies are more closely associated with lower reasoning performance than background characteristics alone. Exploratory clustering further indicates that AI users do not form a single homogeneous group, but instead reflect tentative behavioral profiles, including over-reliant users, mixed-strategy users, and balanced support-seekers. Although the findings are exploratory, they indicate that AI does not affect critical thinking in a uniformly negative or positive way. Instead, its influence appears to depend on the manner in which it is used. The paper therefore argues that effective human--AI collaboration should support reflection, verification, and sustained cognitive effort rather than substitute for them.
\end{abstract}


\section{Introduction}
Artificial Intelligence (AI) has quickly become part of everyday life. Students use it to summarize lessons and understand difficult topics, professionals rely on it to draft reports and solve technical problems, and general users increasingly turn to it for quick answers, explanations, and decision support \cite{unesco_genai_2023}. It is also increasingly used for preliminary health-related guidance when individuals feel unwell, as well as for generating ideas in academic, professional, and creative contexts. Because of this widespread accessibility, AI is no longer limited to specialized technical settings. It has gradually become an everyday cognitive companion that shapes how people approach learning, reasoning, and problem-solving \cite{odea_2024,gonsalves_2025}..

This growing integration of AI into routine thinking has also raised an important concern. Does frequent use of AI strengthen human thinking by supporting it, or does it weaken independent reasoning by encouraging dependence? In many situations, AI clearly saves time and effort, and this can be beneficial. When people repeatedly depend on AI for tasks they could otherwise try to solve on their own, they may begin to reflect less, evaluate less carefully, and lose patience with difficult problems. This pattern is closely related to cognitive offloading, where external tools reduce the need for internal mental effort \cite{risko_gilbert_2016}. Although offloading can improve efficiency, excessive dependence may come at the cost of critical thinking.

Critical thinking is not simply a matter of arriving at the correct answer. It also involves interpretation, logical evaluation, independent judgment, and the persistence needed to work through uncertainty \cite{facione_1990}. These qualities have become even more important in an era where AI systems can produce responses that are fluent and convincing, but sometimes inaccurate, incomplete, or misleading \cite{unesco_genai_2023}. When users accept AI-generated outputs without questioning assumptions or checking alternatives, their intellectual autonomy may gradually weaken \cite{cabellos_2024, essien_2024}.

Motivated by this concern, the present study examines the relationship between AI-use behavior and critical-thinking performance through a structured survey-based assessment. Rather than depending only on self-reported opinions, the study combines participants' perceptions of AI use with objective performance on short logic and reasoning tasks. In this way, it attempts to capture both how individuals believe AI affects them and how they actually perform when asked to reason without assistance.

The main contributions of this study are threefold. First, it documents descriptive patterns of AI-use behavior within a young participant group that is already familiar with regular AI-assisted problem-solving. Second, it connects these behavioral patterns to a Critical Thinking Score (CTS) derived from objective reasoning items rather than relying only on subjective perceptions. Third, it applies exploratory machine learning techniques to identify broad behavioral user profiles and to estimate which self-reported factors are most closely associated with reasoning performance.

Because this study relied on one-to-one interview-administered data collection, the process was both time-consuming and difficult to arrange, which limited the size of the final sample. For this reason, the machine learning analysis is interpreted as exploratory pattern discovery rather than strong predictive modeling. Overall, the paper seeks to answer a simple but timely question: in the age of AI, what kinds of usage habits are most compatible with preserving independent human reasoning? \cite{frederick_2005,risko_gilbert_2016}

\section{Related Work}
Critical thinking is generally understood as purposeful, self-regulatory judgment involving interpretation, analysis, evaluation, and inference \cite{facione_1990}. It has also been described as a reflective stance toward evidence, reasons, and decisions about what to believe or do, rather than as a simple function of intelligence or subject knowledge \cite{ennis_1985}. A closely related concept is cognitive reflection, which refers to the tendency to resist an immediate intuitive answer and engage in further reasoning. For this reason, short reasoning tasks inspired by the Cognitive Reflection Test are often used as compact indicators of reflective and analytic engagement, although such tasks provide only a partial view of broader critical-thinking ability \cite{frederick_2005}.

A key theoretical lens for understanding the relationship between AI use and reasoning is cognitive offloading. This refers to the transfer of mental effort to external tools, often in exchange for reduced time, lower effort, or greater convenience \cite{risko_gilbert_2016}. Earlier studies on digital information access showed that when people expect information to remain available externally, they may remember less of the content itself and more about where to retrieve it \cite{sparrow_liu_wegner_2011}. Other perspectives argue that external tools can sometimes operate as extensions of cognition rather than simple substitutes, which means that offloading is not automatically harmful \cite{clark_chalmers_1998}. However, research on automation has long warned that convenient and seemingly authoritative systems may encourage misuse, disuse, or poorly calibrated trust, especially when users reduce monitoring and independent judgment \cite{parasuraman_riley_1997,lee_see_2004}. This concern becomes even more relevant in the case of generative AI, where fluent language can create a misleading impression of reliability.

Recent empirical work on generative AI presents a mixed but informative picture. Some studies report clear benefits. For example, a large meta-analysis found positive average effects of ChatGPT on learning performance, learning perception, and higher-order thinking, while also noting that these effects vary by instructional role, course type, and duration \cite{wang_fan_2025}. At the same time, not all positive findings necessarily imply stronger independent reasoning. In many cases, the measured gains are tied to assisted performance rather than unaided thinking. This limitation is important because other evidence suggests that unrestricted AI assistance can weaken long-term retention by reducing the effortful processing needed for durable learning \cite{barcaui_2025}. Behavioral studies also show that users may overrely on AI-generated advice even when it conflicts with available evidence or their own judgment \cite{klingbeil_etal_2024}. Taken together, these studies suggest that generative AI can support learning and performance, but they also show that convenience may come at the cost of verification, persistence, and independent reasoning when AI is used as a substitute rather than as a support.

Another important line of research shows that AI users are not a uniform group. Recent studies suggest that stronger critical thinking is associated with more reflective and cautious patterns of AI reliance, whereas trust without sufficient evaluation is linked to more thoughtless use \cite{hou_etal_2025}. Related work on dependency has introduced measures of withdrawal-like discomfort, negative consequences, and cognitive preoccupation, showing that dependence-oriented AI use may be associated with lower task performance and weaker critical-thinking-related outcomes \cite{goh_etal_2025}. Person-centered profiling studies similarly indicate that users differ in their motivations, trust, and perceived value of AI tools \cite{gerling_etal_2026}. However, much of this literature remains limited by cross-sectional self-report designs, attitude-based measures, or outcomes that are not based on direct unaided reasoning performance. Survey-based evidence has also linked frequent AI use and cognitive offloading with weaker critical-thinking-related outcomes, but such studies often rely heavily on perceived behavior rather than objective task performance \cite{gerlich_2025}. 

\subsection{Research Gap}
Overall, the existing literature shows that generative AI is not simply good or bad for human thinking. In some settings, it improves efficiency and supports learning, while in others it encourages overreliance and reduces independent effort. Much of the prior work, however, has focused on perceived learning gains, trust, or task-specific behaviors. Comparatively fewer studies combine everyday AI-use habits with an objective measure of unaided reasoning, and even fewer attempt to profile different types of AI users in relation to critical-thinking performance. This leaves an important gap in understanding how dependence-related behaviors, short-form reasoning ability, and user-level behavioral patterns interact. The present study addresses this gap by combining survey-based measures of AI-use habits with a reasoning-based Critical Thinking Score (CTS) and exploratory profiling methods to examine which styles of AI use are more compatible with preserving independent critical thinking \cite{gerlich_2025,gerling_etal_2026}.

\section{Methodology}

\subsection{Study Design and Participants}
This study adopted a cross-sectional survey design with an embedded reasoning assessment. Data were collected through one-to-one interview-administered sessions rather than a fully self-administered form. In each session, the researcher presented the questions in order, clarified wording when necessary, and recorded responses in a structured format. This approach was chosen to reduce misunderstanding and ensure consistent completion of the instrument. The study focused on two related dimensions: participants' behavioral and attitudinal relationship with AI tools, and their performance on short unaided reasoning tasks.

The participant group consisted of 22 respondents, most of whom were students, with a small number of freelancers or self-employed individuals and one professional participant. They came from different academic backgrounds, with Science/Engineering and Business/Economics being the most common. Most had already been using AI tools regularly for at least several months, and many reported one to two years or more of experience. Because the sample was relatively small and based on convenience access, the study does not claim population-level generalization, but instead aims to identify preliminary patterns that may inform future larger-scale research.

\subsection{Instrument and Scoring}

The instrument contained two parts. The first collected background information and self-reported perceptions of AI use, including age, education, occupation, years of AI usage, commonly used tools, and primary reasons for use. It also included Likert-scale items on confidence before and after regular AI use, perceived change in critical thinking, independence without AI, whether AI deepens or replaces thinking, discomfort without AI, reduced patience in problem-solving, and perceived efficiency. Most of these items were measured on a 5-point Likert scale, where higher values indicated stronger agreement \cite{likert_1932}. Table~\ref{tab:instrument_examples} shows a few illustrative examples of the question types included in the instrument.

\begin{table}[!t]
\caption{Illustrative Instrument Items}
\label{tab:instrument_examples}
\centering
\begin{tabular}{p{1.55cm} p{4.7cm}}
\toprule
\textbf{Type} & \textbf{Example} \\
\midrule
Background & Age, education, occupation, years of AI use \\
Behavioral & ``I can solve most problems without AI if I try.'' \\
Behavioral & ``AI has reduced my patience to struggle with problems.'' \\
Reasoning & ``A bat and a ball together cost \$1.10...'' \\
Reasoning & ``If 10 workers need 10 days, how many days will 5 workers need?'' \\
Reflection & ``Is AI strengthening or weakening human thinking? Why?'' \\
\bottomrule
\end{tabular}
\end{table}

The second part consisted of seven short reasoning questions designed to assess independent analytical ability. These tasks covered cognitive reflection, numerical reasoning, proportional reasoning, algebraic reasoning, temporal reasoning, and productivity logic. Representative examples included the bat-and-ball problem, geometric sequence completion, worker-task proportionality, algebraic inversion, pill-timing reasoning, travel-speed calculation, and machine-productivity logic \cite{frederick_2005}. Participants were also asked whether they had used AI while answering these tasks and, if so, what role it played.

Each reasoning item was manually coded as correct or incorrect. A \textit{Critical Thinking Score} (CTS) was then computed as
\[
\mathrm{CTS} = \frac{\text{Number of Correct Answers}}{7}\times 100.
\]
This percentage score served as the main outcome variable in the analysis \cite{facione_1990,frederick_2005}. The behavioral variables were retained as explanatory features, and Likert-scale responses were encoded numerically so that higher values consistently reflected stronger agreement. In particular, higher values on discomfort without AI and reduced patience were interpreted as stronger indicators of AI dependence or reduced cognitive persistence.

\subsection{Analytical Procedure}
The analysis was conducted in three stages. First, descriptive statistics were used to summarize participant characteristics, AI-use behavior, and overall reasoning performance. Second, correlation analysis was applied to examine associations between selected behavioral variables and CTS. Third, exploratory machine learning methods were used to identify behavioral profiles and estimate the relative importance of self-reported features. Specifically, K-Means clustering was used to identify broad user groups based on dependency-related behavioral variables \cite{macqueen_1967}, Random Forest was used to estimate feature importance for variation in CTS \cite{breiman_2001}, and Principal Component Analysis (PCA) was used to visualize the clusters in a lower-dimensional space \cite{jolliffe_2002}. Because the sample size was limited, these machine learning results were interpreted as exploratory pattern discovery rather than confirmatory evidence.

\section{Results and Discussion}

\subsection{Overall Survey Outcomes}
Table~\ref{tab:sample_summary} summarizes the main characteristics of the sample. Most respondents were students in the 21--23 age range, and most had already been using AI tools regularly for at least several months. ChatGPT was the most commonly reported tool, often used alongside Gemini, Grok, DeepSeek, or Copilot. The dominant motivations were learning concepts, solving problems, generating ideas, and saving time. When facing difficult problems, most participants either tried to solve them first or thought briefly before turning to AI, while only a small minority reported asking AI immediately.

\begin{figure}[htbp]
\centering
\includegraphics[width=0.48\textwidth]{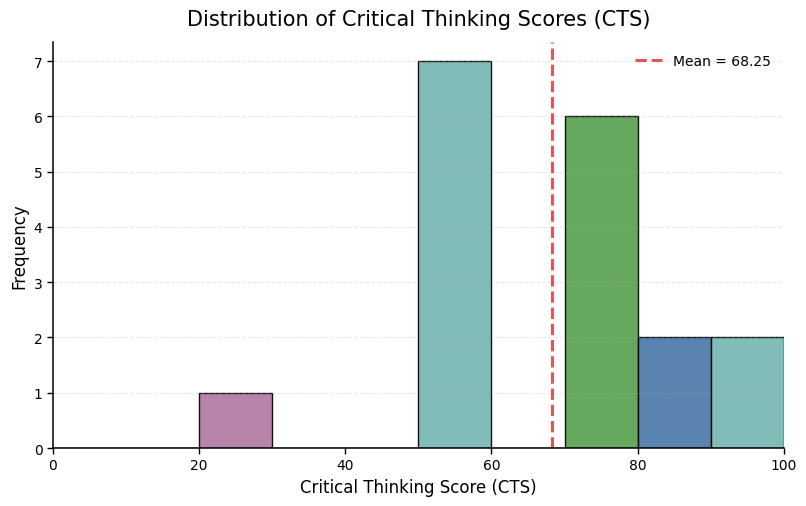}
\caption{Distribution of Critical Thinking Scores (CTS) among the participants. The dashed vertical line indicates the sample mean, while the spread of the histogram shows substantial variation in unaided reasoning performance across respondents.}
\label{fig:cts_distribution}
\end{figure}

Self-perceived confidence generally increased after regular AI use, but this did not necessarily imply preserved independent reasoning. Many participants agreed that AI helps them think deeper, yet a substantial number also reported discomfort without AI and reduced patience when struggling with difficult problems. Most respondents completed the reasoning section without declared AI assistance. The mean Critical Thinking Score (CTS) was 68.25\%, corresponding to an average of approximately 4.78 correct answers out of 7, and the score distribution showed noticeable variability across participants, as illustrated in Fig.~\ref{fig:cts_distribution}.

\begin{table}[htbp]
\caption{Descriptive Summary of the Sample ($N=22$)}
\label{tab:sample_summary}
\centering
\begin{tabular}{>{\raggedright\arraybackslash}p{4.15cm}c}
\toprule
\textbf{Variable} & \textbf{Value} \\
\midrule
Participants & 22 \\
Age 21--23 & 50.0\% \\
Students & 86.4\% \\
Science/Engineering background & 54.5\% \\
Regular AI usage 1--2 years & 36.4\% \\
Regular AI usage $>$ 2 years & 31.8\% \\
Reported ChatGPT use & 100.0\% \\
Try solving first before AI & 40.9\% \\
Think briefly, then use AI & 40.9\% \\
Immediately ask AI & 13.6\% \\
Reported no AI use in Section B & 90.9\% \\
Mean Critical Thinking Score & 68.25\% \\
Average correct answers & 4.78 / 7 \\
\bottomrule
\end{tabular}
\end{table}

\subsection{Item-Level Performance on Reasoning Tasks}
\begin{figure}[htbp]
\centering
\includegraphics[width=0.48\textwidth]{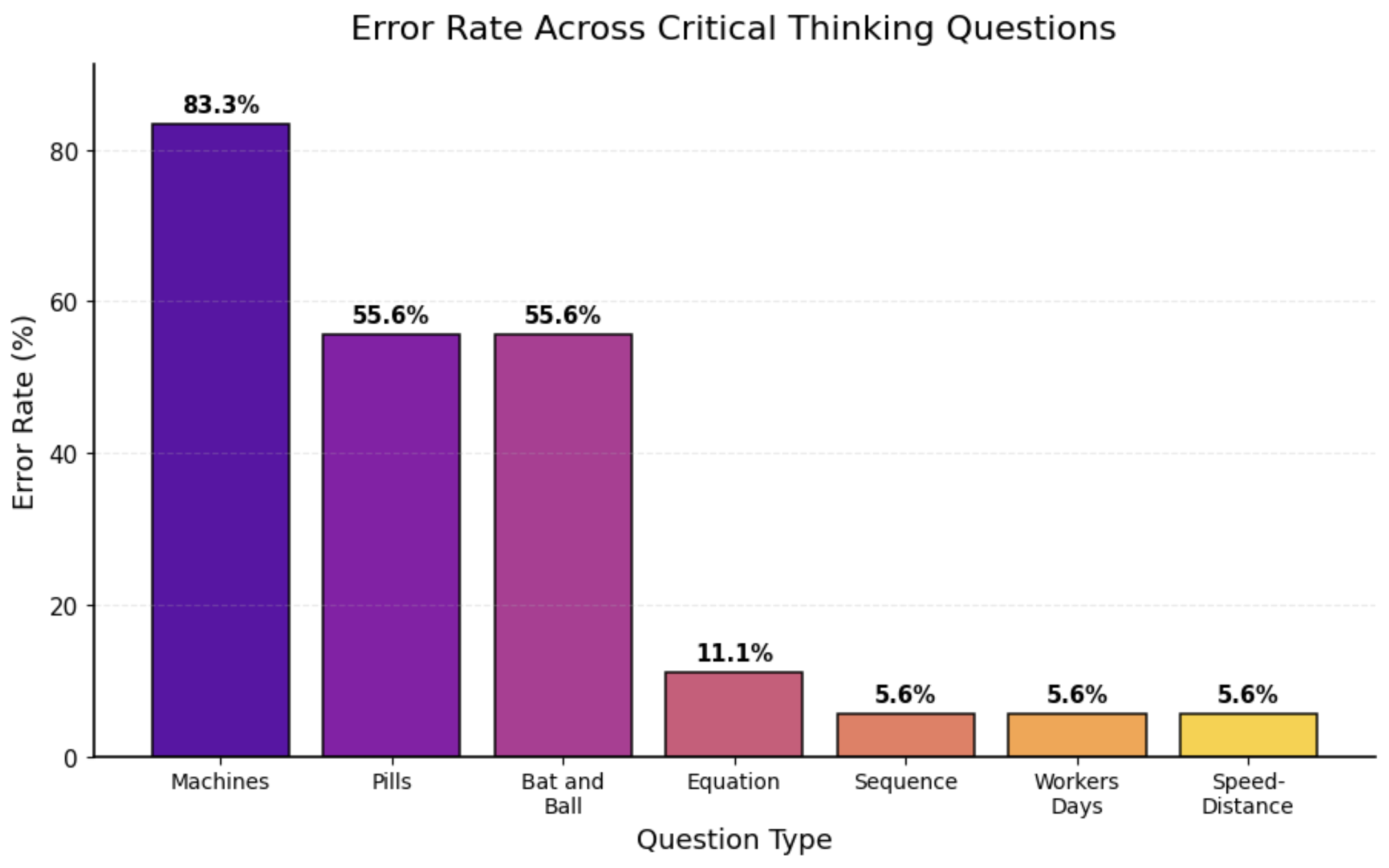}
\caption{Error rates across the seven critical-thinking questions. The machine-productivity item produced the highest error rate, while sequence, workers-days, and speed-distance showed the lowest error rates. Higher values indicate questions that were more difficult for participants to answer correctly.}
\label{fig:error_rate}
\end{figure}

To examine which reasoning tasks were most difficult, error rates were computed for each of the seven questions. As shown in Fig.~\ref{fig:error_rate}, the machine-productivity item had the highest error rate, followed by the pill-timing and bat-and-ball problems, whereas the sequence, workers-days, and speed-distance items were answered more accurately. This suggests that participants handled direct procedural reasoning more successfully than questions requiring cognitive reflection or resistance to intuitive but incorrect responses \cite{frederick_2005}.

\subsection{Behavioral Patterns and Correlation Insights}
Overall, the survey reveals a clear tension in AI use. Many participants viewed AI as a helpful tool for productivity and learning, but several responses also suggested that convenience may weaken cognitive endurance. In particular, respondents often indicated that AI can be beneficial when used for hints or explanation, but harmful when it becomes a substitute for thinking.

\begin{figure*}[htbp]
\centering
\includegraphics[width=0.92\textwidth]{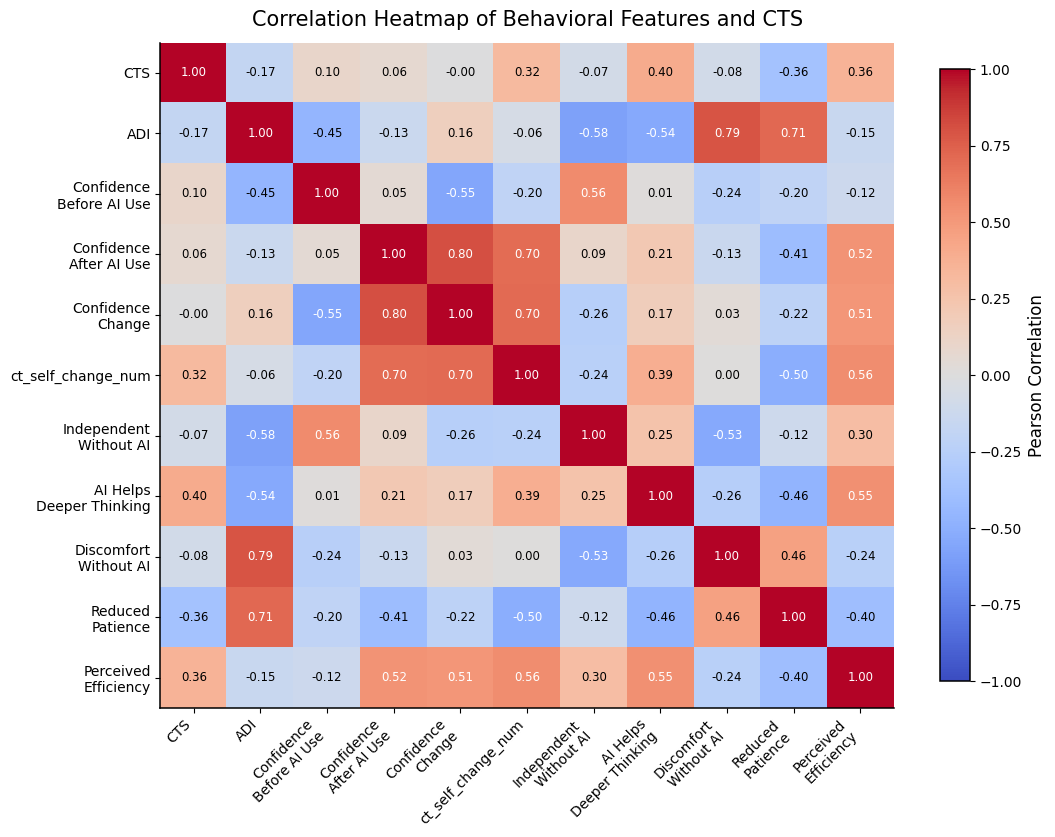}
\caption{Correlation heatmap of selected behavioral features and Critical Thinking Score (CTS). Negative values indicate inverse relationships with reasoning performance, while positive values indicate direct associations among behavioral variables.}
\label{fig:corr_heatmap}
\end{figure*}

This pattern is supported by the correlation analysis. CTS showed a moderate negative relationship with the statement \textit{``AI has reduced my patience to struggle with problems''} ($r=-0.36$), indicating that participants who reported greater patience loss tended to perform worse on the reasoning tasks. By contrast, perceived efficiency did not show a similarly strong positive relationship with reasoning performance. Together, these findings suggest that habit formation may matter more than confidence alone, and that reduced willingness to engage with difficulty may be more important than simply feeling that AI is useful. The broader correlation structure is shown in Fig.~\ref{fig:corr_heatmap} \cite{hiebert_grouws_2007}.

\subsection{User Segmentation via K-Means}
K-Means clustering was applied to identify broad behavioral profiles among participants. The results suggested three tentative groups. The first cluster reflected more over-reliant users, who showed higher discomfort without AI and lower patience, and also had the lowest mean CTS. The second cluster represented mixed-strategy users, who achieved the highest average CTS but also showed the greatest variability. The third cluster reflected more balanced support-seekers, who appeared to use AI after some initial effort rather than as an immediate replacement for thought. Although these groups are exploratory rather than definitive, they suggest that AI users are not homogeneous and that usage style may shape reasoning outcomes. Figure~\ref{fig:pca_profiles} presents the PCA-based visualization of these behavioral profiles.

\begin{figure}[htbp]
\centering
\includegraphics[width=0.48\textwidth]{pca_plot.png}
\caption{PCA projection of the K-Means behavioral user profiles. The three clusters were tentatively interpreted as Over-Reliant Users, Mixed-Strategy Users, and Balanced Support-Seekers based on their relative patterns of AI dependency, patience loss, and independent reasoning performance. The X markers indicate cluster centroids in the two-dimensional PCA space.}
\label{fig:pca_profiles}
\end{figure}

\subsection{Exploratory ML: Feature Importance}
\begin{figure}[htbp]
\centering
\includegraphics[width=0.48\textwidth]{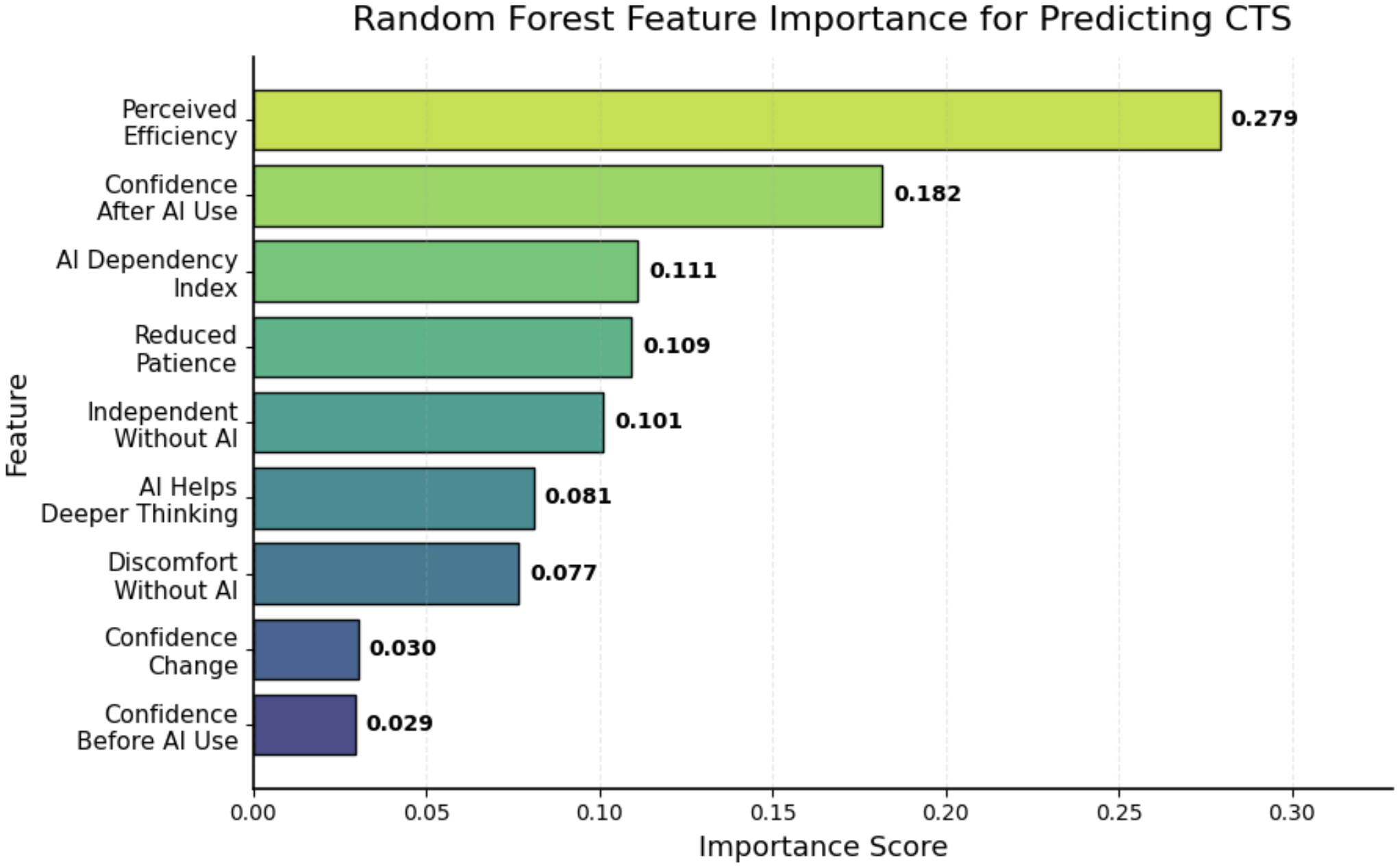}
\caption{Random Forest feature importance scores for predicting Critical Thinking Score (CTS). Perceived efficiency and confidence after AI use received the highest relative importance in this exploratory model, while reduced patience and AI dependency also contributed meaningfully.}
\label{fig:rf_importance}
\end{figure}

Random Forest analysis was used to estimate which behavioral variables were most strongly associated with variation in CTS. The results indicated that perceived efficiency and confidence after AI use received the highest relative importance in the exploratory model, while reduced patience and AI dependency also contributed meaningfully. This suggests that perceived gains in efficiency and confidence do not necessarily translate into stronger independent reasoning, especially when accompanied by reduced persistence. Because the sample is small, these importance values should not be interpreted as definitive, but they remain useful for ranking the relative influence of variables within this dataset. Figure~\ref{fig:rf_importance} summarizes the feature-importance results.

\subsection{Discussion}
Taken together, the findings suggest that AI does not influence critical thinking in a single fixed way. Its effect appears to depend on the habits of use it encourages. Participants who used AI while preserving effort, reflection, and verification did not necessarily perform poorly, whereas those reporting reduced patience and stronger discomfort without AI tended to show lower reasoning performance. This implies that the central issue is not whether AI should be used, but whether it is used as a support for thinking or as a substitute for it. In educational and professional settings, this supports the value of human-in-the-loop strategies, such as requiring an initial attempt before consulting AI, encouraging users to justify AI-generated answers in their own words, and designing systems that promote reflection rather than passive acceptance \cite{lee_2024,unesco_genai_2023,hiebert_grouws_2007}.

\subsection{Limitations}
This study has several limitations. The sample size is small, the participant group is dominated by students, and several variables rely on self-reported perceptions. In addition, the reasoning items provide only a limited snapshot of critical thinking rather than a comprehensive assessment of real-world analytical ability. For these reasons, the machine learning results should be treated as exploratory. Future work should use larger and more diverse samples, richer critical-thinking instruments, and possibly longitudinal designs to examine whether these behavioral patterns persist over time.

\section{Conclusion}
This study examined the relationship between AI-use behavior and critical-thinking performance through an interview-administered survey and a short reasoning assessment. The findings suggest that AI does not affect human thinking in a uniformly positive or negative way; rather, its influence appears to depend on how it is used and whether users remain cognitively engaged while relying on it. The results indicate that reduced patience for struggling with difficult problems may be a more important warning sign than AI use alone, that AI users reflect different behavioral profiles rather than a single homogeneous group, and that perceived efficiency or increased confidence after AI use should not be automatically interpreted as stronger independent reasoning. Overall, the study suggests that the key challenge in the AI era is not whether AI should be used, but how it should be used. When AI supports reflection, feedback, and explanation, it may complement human reasoning; when it replaces effort, verification, and productive struggle, it may weaken important habits of independent thought. Educational and technological systems should therefore encourage forms of human--AI collaboration that preserve reflection, verification, and sustained cognitive effort rather than reduce them.

\bibliographystyle{IEEEtran}
\bibliography{references}

\end{document}